\documentclass[a4paper,conference]{IEEEtran}

\usepackage{amsmath}

\usepackage{cite}      

\usepackage{graphicx}  

\usepackage{subfigure} 
\usepackage{aas_macros}
\usepackage{fancyheadings}
\begin{document}

\title{An Integral-based Approach for the Vector Potential in Smoothed Particle Magnetohydrodynamics}

\author{\IEEEauthorblockN{Terrence S. Tricco}
\IEEEauthorblockA{Department of Computer Science\\
Memorial University of Newfoundland\\
St. John's, NL, Canada\\
tstricco@mun.ca}
\and
\IEEEauthorblockN{Daniel J. Price}
\IEEEauthorblockA{School of Physics and Astronomy\\
Monash University\\
Clayon, VIC, Australia\\
daniel.price@monash.edu}}


\maketitle

\begin{abstract}
A new implementation for the time evolution of the magnetic vector potential is obtained for smoothed particle magnetohydrodynamics by considering the induction equation in integral form. Galilean invariance is achieved through proper gauge choice. This new discretisation is tested using the Orszag-Tang MHD vortex in a 3D configuration. The corresponding conservative equations of motion are derived, but are not found to solve the MHD equations in the continuum limit. Tests are performed using a hybrid approach instead, whereby the equations of motion based on the magnetic field instead of vector potential are used. Test results experience the same numerical instability as with the Price (2010) formulation. We conclude that this new formulation is non-viable.
\end{abstract}

\section{Introduction}

Smoothed particle magnetohydrodynamics (SPMHD) has grown to become a robust method for solving the equations of magnetohydrodynamics, that is, of magnetised fluids. It has been shown that SPMHD can capture the magneto-rotational instability \cite{wissing-etal22} and the small-scale dynamo amplification of magnetic energy in magnetised turbulence \cite{tpf16}. It has been used to study the magnetic field structure in a wide of array of problems, such as star formation \cite{wpb16, wbp19}, the Milky Way \cite{dobbs-etal16}, tidal disruption events \cite{bonnerot-etal17}, and magnetically confined plasmas \cite{vvsrb19}.

There were many pieces of the SPMHD puzzle that had to be solved to reach this point. Some of the key steps were determining how to handle instability caused by magnetic tension \cite{bot01}, how to formulate dissipative terms \cite{pm05}, and upholding the divergence-free constraint of the magnetic field \cite{tp12, tpb16}. Some of these problems are linked. From Maxwell's equations, $\nabla \cdot{\bf B} = 0$, where ${\bf B}$ is the magnetic field. Numerical errors can lead to non-zero divergence errors, which in turn can grow in time leading to unphysical behaviour. Constrained hyperbolic divergence cleaning \cite{tp12, tpb16} is a numerically stable method that typically keeps the average divergence error to around 1\%, as measured by the mean $h \vert \nabla \cdot {\bf B} \vert / \vert {\bf B} \vert$.

While divergence cleaning has been successful in the context of SPMHD, there is one substantive shortcoming -- it is an approximate method for upholding the divergence-free constraint. Ideally, the divergence-free condition would be exactly upheld by the numerical method. One approach is to formulate the magnetic field in terms of the vector potential, that is, ${\bf B} = \nabla \times {\bf A}$, which guarantees a divergence-free reconstruction because the divergence of the curl is zero. An SPMHD formulation in terms of the vector potential was derived by \cite{price10}, henceforth P10, but found that it suffered from numerical instability. The time evolution of the vector potential led to exponential growth of total energy, and the conservative equations of motion derived using the vector potential were also found to be numerically unstable.

In this paper, a different formulation for the time evolution of the vector potential is proposed. Past approaches have written the induction equation in terms of the vector potential and discretised the resultant equation \cite{price10, se15}. Instead, for this work, the discretised equation is derived by considering the induction equation using a volume integral approach. This leads to a novel discretisation.

Section~\ref{sec:vp} provides essential background. It introduces the continuum induction equation written in terms of the vector potential, how to reconstruct the magnetic field from the vector potential in SPHMHD, and the P10 formulation \cite{price10}. Section~\ref{sec:volume} derives the new formulation based on a volume integral approach. Section~\ref{sec:momentum} discusses the equations of motion, and, in particular, provides the corresponding equations of motion for the new formulation. Test results are presented in Section~\ref{sec:tests} for the 3D Orszag-Tang vortex, and conclusions are given in Section~\ref{sec:conclusions}.

\section{Vector Potential}
\label{sec:vp}

The induction equation for the time evolution of the magnetic field is
\begin{equation}
\label{eq:dbdt}
\frac{\partial {\bf B}}{\partial t} = \nabla \times  \left({\bf v} \times {\bf B} \right) ,
\end{equation}
where ${\bf v}$ is the velocity. In standard SPMHD, the induction equation is written in terms of a Lagrangian derivative, and the right-hand side is expanded with the term proportional to $\nabla \cdot {\bf B}$ removed, leading to
\begin{equation}
\frac{{\rm d}{\bf B}}{{\rm d}t} = - ({\bf B} \cdot \nabla) {\bf v} + {\bf B} (\nabla \cdot {\bf v}).
\end{equation}

To create an equation which evolves ${\bf A}$ forward in time, ${\bf B} = \nabla \times {\bf A}$ is first substituted into (\ref{eq:dbdt}), yielding
\begin{equation}
\label{ed:dadt}
\frac{\partial (\nabla \times {\bf A})}{\partial t} = \nabla \times \left({\bf v} \times {\bf B} \right) .
\end{equation}
which can be `un-curled' to yield
\begin{equation}
\label{eq:dadt2}
\frac{\partial {\bf A}}{\partial t} = {\bf v} \times {\bf B} + \nabla \phi ,
\end{equation}
where $\nabla \phi$ is a constant of integration representing the gauge choice. In the Lagrangian frame, this becomes
\begin{equation}
\label{eq:dadt3}
\frac{{\rm d} {\bf A}}{{\rm d} t} = {\bf v} \times {\bf B} + \left( {\bf v} \cdot \nabla \right) {\bf A} + \nabla \phi .
\end{equation}
The $({\bf v} \cdot \nabla) {\bf A}$ term is a `reverse advection' term added to obtain the Lagrangian time derivative on the left-hand side.

\subsection{Reconstructing {\bf B} in SPMHD}

There is freedom to choose how to estimate ${\bf B} = \nabla \times {\bf A}$ in SPMHD. A logical choice is to use a difference derivative, given by
\begin{equation}
\label{eq:b}
{\bf B}_a = \frac{1}{\Omega_a \rho_a} \sum_b m_b \left( {\bf A}_a - {\bf A}_b \right) \times \nabla_a W_{ab}(h_a) ,
\end{equation}
where $\Omega$ accounts for gradients in the smoothing length, $\rho$ is the density, $W$ is the smoothing kernel, and $h$ is the smoothing length.

\subsection{Price (2010) Formulation}

The P10 formulation used a gauge choice of $\phi = - {\bf v} \cdot {\bf A}$. By using the identity
\begin{align}
\label{eq:gauge_identity}
\nabla ({\bf v} \cdot {\bf A}) =&~ ({\bf v} \cdot \nabla) {\bf A} + {\bf v} \times (\nabla \times {\bf A}) \nonumber \\
&+ ({\bf A} \cdot \nabla) {\bf v} + {\bf A} \times (\nabla \times {\bf v}),
\end{align}
(\ref{eq:dadt3}) will simplify to
\begin{equation}
\label{eq:p10}
\frac{{\rm d}{\bf A}}{{\rm d}t} = - {\bf A} \times (\nabla \times {\bf v}) - \left( {\bf A} \cdot \nabla \right) {\bf v} .
\end{equation}
This can equivalently be expressed in tensor notation as 
\begin{equation}
\frac{{\rm d}A^i}{{\rm d}t} = -  A_j \frac{\partial v^j}{\partial x^i} .
\end{equation}
The benefit of this gauge choice is that the derivatives are moved onto the velocity. When discretised, this leads to a Galilean invariant form, given by
\begin{equation}
\label{eq:p10}
\frac{{\rm d}A_a^i}{{\rm d}t} = \frac{{A_a^j}}{\Omega_a \rho_a} \sum_b m_b \left( v_a^j - v_b^j \right) \frac{\partial W_{ab}(h_a)}{\partial x_a^i} .
\end{equation}
However, \cite{price10} found that this implementation suffers from numerical instability. Even worse, the corresponding conservative equations of motion for this discretisation are also numerically unstable.

\section{Volume Integral Discretisation}
\label{sec:volume}

Taking inspiration from \cite{mocz-etal16}, (\ref{eq:dadt2}) could instead be written in integral form according to
\begin{equation}
\frac{{\rm d}}{{\rm d}t} \int_V {\bf A} {\rm d}V = \int_V {\bf v} \times {\bf B} {\rm d}V + \int_{\partial V} {\bf A} {\bf v} \cdot {\rm d}S,
\end{equation}
where the second term represents the advection of a surface. Note that the time derivative has changed from Eulerian to Lagrangian derivative.

In SPMHD, there are no well-defined surfaces. Instead, the surface integral is converted to a volume integral (which is well defined), obtaining
\begin{equation}
\label{eq:dadt4}
\frac{{\rm d}}{{\rm d}t} \int_V {\bf A} {\rm d}V = \int_V {\bf v} \times {\bf B} {\rm d}V + \int_V \nabla_j \left({\bf A} {\bf v}^j  \right) {\rm d}V.
\end{equation}
The integral element is taken to be $\rho {\rm d}V$, which corresponds to the mass element. Equation~(\ref{eq:dadt4}) is discretised to yield
\begin{equation}
\frac{{\rm d}}{{\rm d}t} \sum_a m_a \frac{{\bf A}_a}{\rho_a} = \sum_a m_a \frac{{\bf v}_a \times {\bf B}_a}{\rho_a} + \sum_a m_a \frac{\nabla_j ({\bf A}_a {\bf v}_a^j)}{\rho_a} .
\end{equation}
This implies that each particle should evolve according to
\begin{equation}
\label{eq:integral_darhodt}
\frac{{\rm d}}{{\rm d}t} \left( \frac{{\bf A}_a}{\rho_a} \right) = \frac{{\bf v}_a \times {\bf B}_a}{\rho _a} + \frac{1}{\rho_a} \nabla_j \left( {\bf A}_a {\bf v}_a^j \right) .
\end{equation}
At this point, there is no specific prescription as to the discretisation used to reconstruct ${\bf B}$ nor for the calculation of the second term. In this work, the difference derivative estimate is chosen for both. That is, the magnetic field is reconstructed according to (\ref{eq:b}) and the second term is calculated using
\begin{align}
\label{eq:second_term}
\frac{1}{\rho_a} \nabla_j ({\bf A}_a {\bf v}_a^j) = &- \frac{1}{\Omega_a \rho_a^2} \sum_b m_b \big[ {\bf A}_a {\bf v}_a \cdot \nabla_a W_{ab}(h_a) \nonumber \\
&\hspace{19mm} - {\bf A}_b {\bf v}_b \cdot \nabla_a W_{ab}(h_a) \big] .
\end{align}

Equation~(\ref{eq:second_term}) can be understood more intuitively through the following analysis. If the term 
\begin{equation}
- \frac{1}{\Omega_a \rho_a^2} \sum_b m_b ({\bf A}_a - {\bf A}_a) {\bf v}_b \cdot \nabla_a W_{ab}(h_a) ,
\end{equation}
which is equivalent to zero, is added to (\ref{eq:second_term}), then
\begin{align}
\label{eq:analysis}
\frac{1}{\rho_a} \nabla_j ({\bf A}_a {\bf v}_a^j) = &- \frac{1}{\Omega_a \rho_a^2} \sum_b m_b ({\bf A}_a - {\bf A}_b) {\bf v}_b \cdot \nabla_a W_{ab}(h_a) \nonumber \\
& - \frac{{\bf A}_a}{\Omega_a \rho_a^2} \sum_b m_b ( {\bf v}_a - {\bf v}_b) \cdot \nabla_a W_{ab}(h_a)
\end{align}
is obtained. The first term in (\ref{eq:analysis}) is interesting because it represents the reverse advection term in (\ref{eq:dadt3}), but uses the weighted summation of the velocity of neighbouring particles instead of the particle's own velocity, that is, ${\bf v}_b$ instead of ${\bf v}_a$. 
The second term in (\ref{eq:analysis}) is equivalent to ${\bf A}_a \nabla \cdot {\bf v}_a$. This arises from the consequence of evolving ${\bf A}/\rho$ instead of just ${\bf A}$, which is evident from 
\begin{equation}
\frac{{\rm d}}{{\rm d}t} \left( \frac{\bf A}{\rho} \right) = \frac{1}{\rho} \frac{{\rm d}{\bf A}}{{\rm d}t} - \frac{{\bf A}}{\rho^2} \frac{{\rm d}\rho}{{\rm d}t} ,
\end{equation}
given that ${\rm d}\rho/{\rm d}t = - \rho \nabla \cdot {\bf v}$. Thus, (\ref{eq:integral_darhodt}) could be stated in terms of evolving ${\bf A}$ directly according to
\begin{equation}
\label{eq:integral_dadt}
\frac{{\rm d}{\bf A}_a}{{\rm d}t} = {\bf v}_a \times {\bf B}_a - \frac{1}{\Omega_a \rho_a} \sum_b m_b ({\bf A}_a - {\bf A}_b) {\bf v}_b \cdot \nabla_a W_{ab}(h_a) .
\end{equation}

\subsection{Galilean Invariance}

Equation~(\ref{eq:integral_dadt}) is not Galilean invariant. However, this can be achieved by the choice of $\phi = - {\bf v} \cdot {\bf A}$ for the gauge. Making use of the identity in (\ref{eq:gauge_identity}), this yields
\begin{align}
\frac{{\rm d}{\bf A}_a}{{\rm d}t} =& - {\bf A}_a \times (\nabla \times {\bf v}_a) - ({\bf A}_a \cdot \nabla) {\bf v}_a \nonumber \\
&+ \frac{1}{\Omega_a \rho_a} \sum_b m_b ({\bf A}_a - {\bf A}_b) \left[ ({\bf v}_a - {\bf v}_b) \cdot \nabla_a W_{ab}(h_a) \right] .
\end{align}
Written in tensor notation, the discretisation is
\begin{align}
\label{eq:integral_dadt_gauge}
\frac{{\rm d}A_a^i}{{\rm d}t} =& \frac{{A_a^j}}{\Omega_a \rho_a} \sum_b m_b \left( v_a^j - v_b^j \right) \frac{\partial W_{ab}(h_a)}{\partial x_a^i} \nonumber \\
& + \frac{1}{\Omega_a \rho_a} \sum_b m_b \left( A_a^i - A_b^i \right) \left[ (v_a^j - v_b^j) \frac{\partial W_{ab}(h_a)}{\partial x_a^j} \right].
\end{align}
It is clear from this that (\ref{eq:integral_dadt_gauge}) represents a novel discretisation of ${\rm d}{\bf A}/{\rm d}t$. By comparison to (\ref{eq:p10}), it is equivalent to the P10 formulation with an additional term. 

\section{Equations of Motion}
\label{sec:momentum}

\subsection{P10 conservative equations of motion}

The conservative equations of motion for the P10 formulation were derived from a Lagrangian variational principle. For simplicity, the equations are given here assuming a constant smoothing length (i.e., no smoothing length gradients), in which case they are
\begin{align}
\frac{{\rm d}v_a^i}{{\rm d}t} =& - \sum_b m_b \left[ \frac{P_a}{\rho_a^2} + \frac{P_b}{\rho_b^2} \right]\frac{\partial W_{ab}}{\partial x_a^i} \nonumber \\
& + \frac{3}{2 \mu_0} \sum_b m_b  \left[ \frac{B_a^2}{\rho_a^2} + \frac{B_b^2}{\rho_b^2} \right] \frac{\partial W_{ab}}{\partial x_a^i} \nonumber \\
& - \frac{1}{\mu_0} \epsilon_{jkl} \sum_b m_b (A_a^k - A_b^k) \left[ \frac{B_a^j}{\rho_a^2} + \frac{B_b^j}{\rho_b^2} \right] \frac{\partial^2 W_{ab}}{\partial x_a^i \partial x_a^l} \nonumber \\
& - \sum_b m_b \left[ \frac{A_a^i}{\rho_a^2} J_a^k + \frac{A_b^i}{\rho_b^2} J_b^k \right] \frac{\partial W_{ab}}{\partial x_a^k} ,
\label{eq:p10-force}
\end{align}
where ${\bf J} = \nabla \times {\bf B} / \mu_0$ is the current density and, in the above, is estimated using a symmetric derivative according to
\begin{equation}
J_a^k = - \frac{\rho_a}{\mu_0} \epsilon_{kjl} \sum_b m_b \left[ \frac{B_a^j}{\Omega_a \rho_a^2} \frac{\partial W_{ab}(h_a)}{\partial x_a^l} + \frac{B_b^k}{\Omega_b \rho_b^2} \frac{\partial W_{ab}(h_b)}{\partial x_a^l} \right] .
\label{eq:symm_J}
\end{equation}

The conservative equations of motion with the P10 formulation were found to be numerically unstable. There are three primary issues. The second term in (\ref{eq:p10-force}) is a negative isotropic pressure, meaning particles will clump when $3 B^2 / 2 \mu_0 > P$. The third term involves the direct second derivative of the kernel, which is known to be noisy. The fourth term includes an estimate of the current density using a symmetric derivative, which is a low-order estimate. Solutions to these problems are not obvious.

\begin{figure*}
{\footnotesize \hspace{1.2in} \sc Volume Integral \hspace{1.05in} Price (2010) \hspace{0.9in} Directly Evolving ${\bf B}$}
\vspace{-3mm}
\begin{center}
\includegraphics[width=2in]{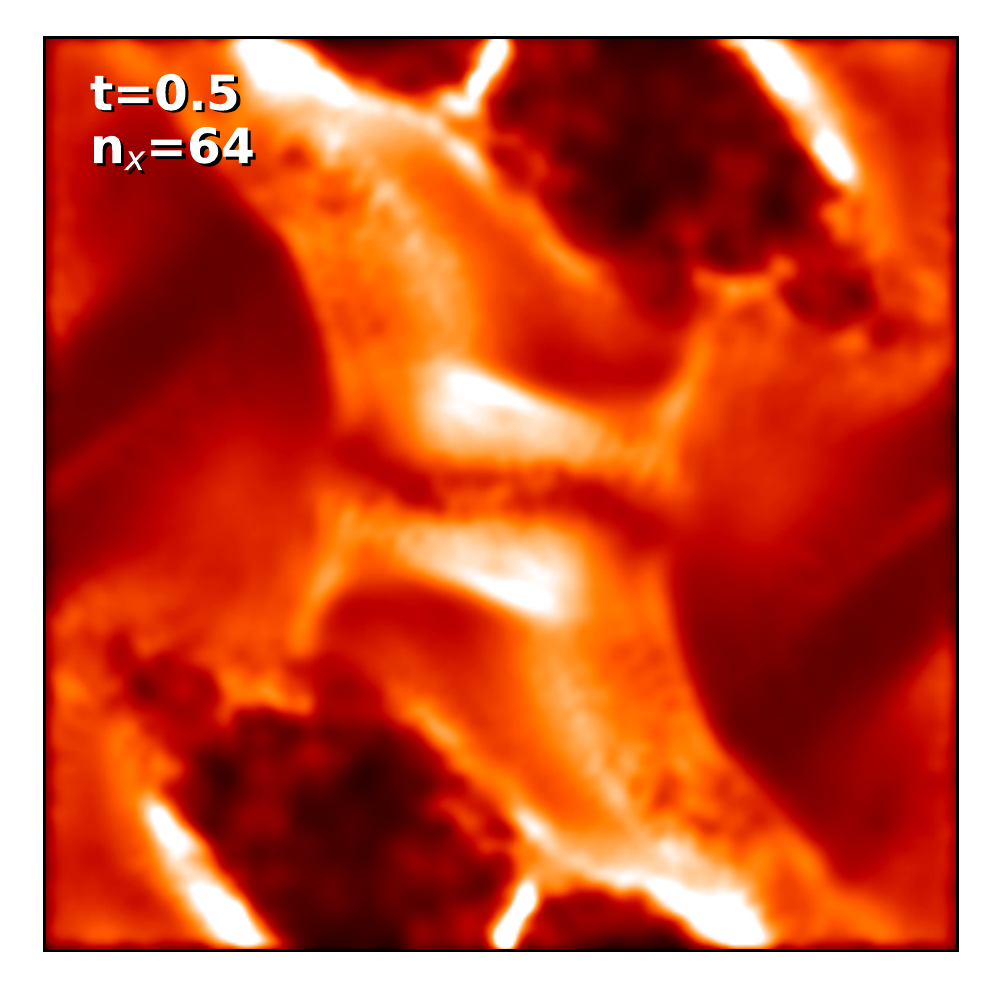}\hspace{-4.5mm}
\includegraphics[width=2in]{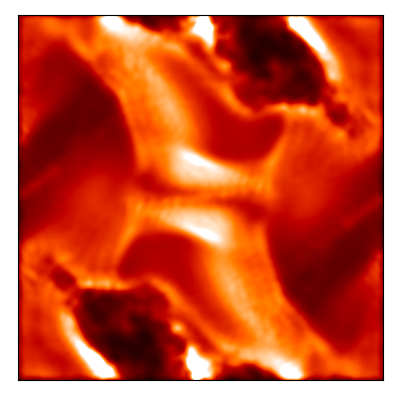}\hspace{-4.5mm}
\includegraphics[width=2in]{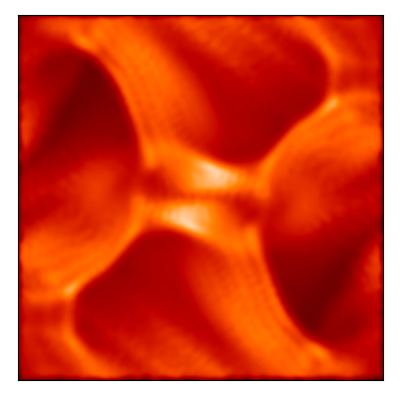} \\
\vspace{-3mm}
\includegraphics[width=2in]{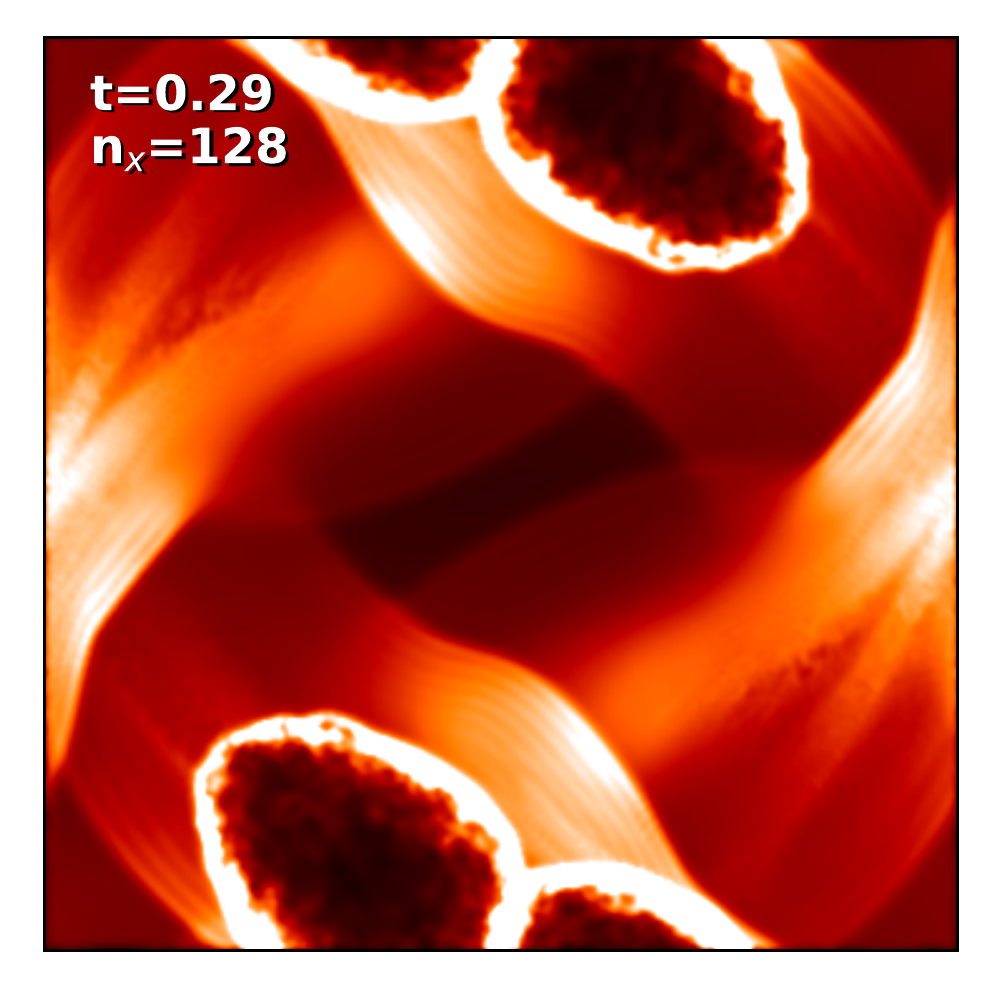}\hspace{-4.5mm}
\includegraphics[width=2in]{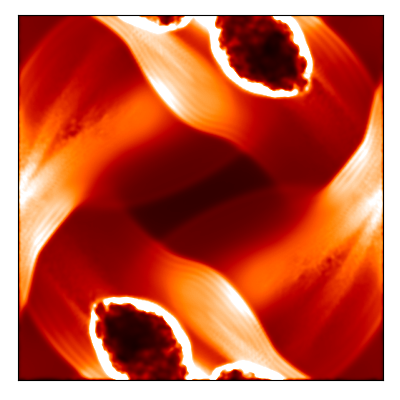}\hspace{-4.5mm}
\includegraphics[width=2in]{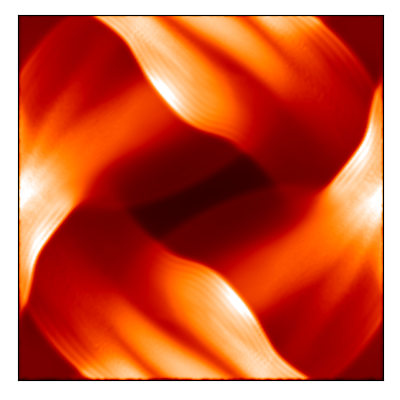} \\
\vspace{-3mm}
\includegraphics[width=2in]{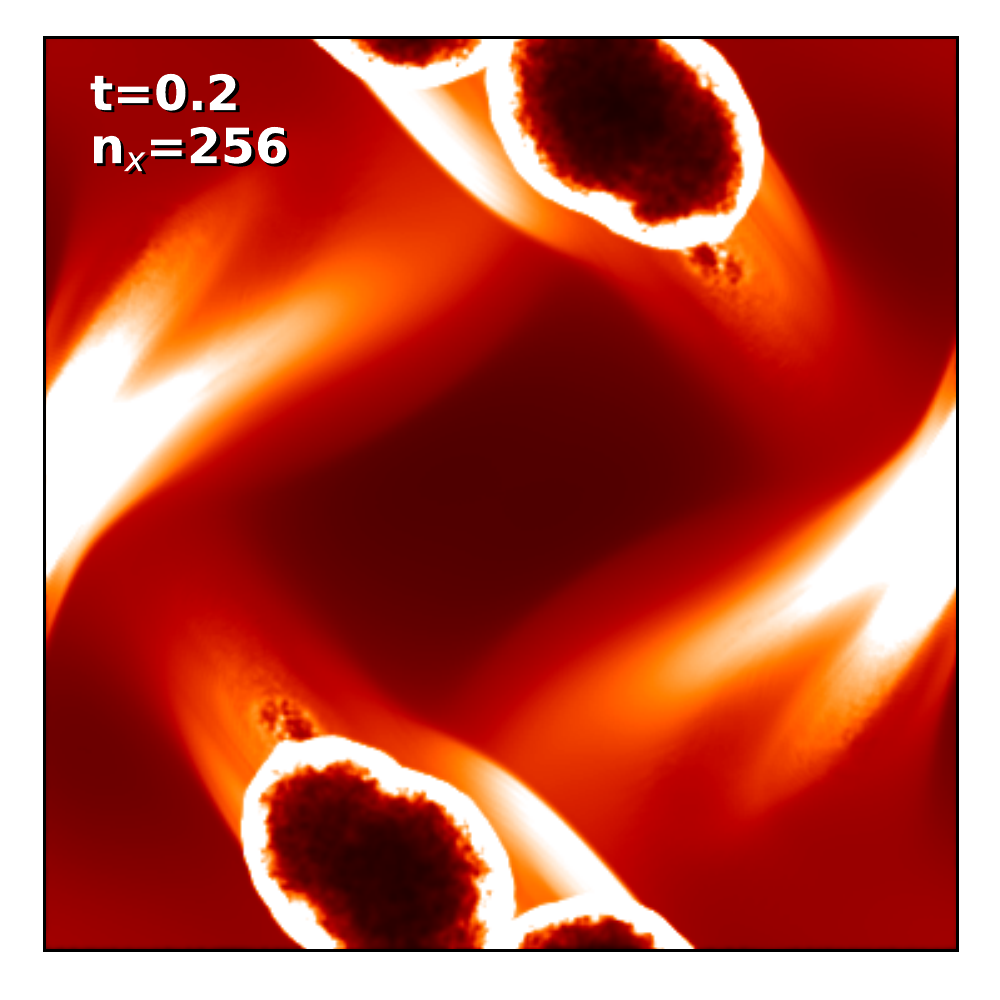}\hspace{-4.5mm}
\includegraphics[width=2in]{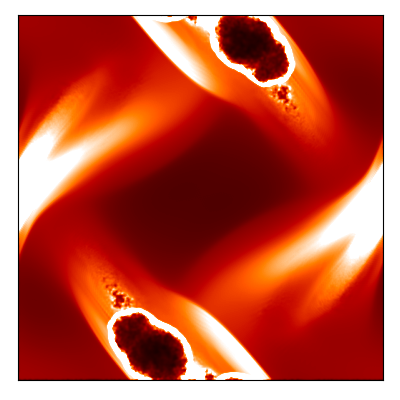}\hspace{-4.5mm}
\includegraphics[width=2in]{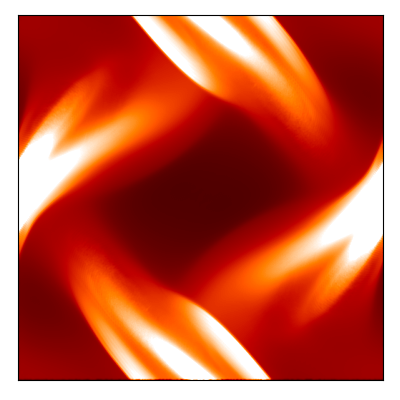}
\end{center}
\caption{Density slices at the midplane. The new vector potential formulation (left column) is compared to the calculations using the P10 formulation (middle column) and directly evolving the magnetic field (right column). Both the volume integral and P10 formulations experience numerical instability at similar times, occurring earlier as the resolution is increased ($t{\sim}0.5$ for $n_{\rm x}=64$ along the top row to $t{\sim}0.19$ for $n_{\rm x}$=256 along the bottom row).}
\label{fig:density}
\end{figure*}

\subsection{Volume integral equations of motion}

The conservative equations of motion for the new volume integral formulation, given by (\ref{eq:integral_dadt_gauge}) using the $\phi = - {\bf v} \cdot {\bf A}$ gauge, can also be derived using a variational principle. 
The full derivation is too lengthy to include here, thus only the final set of equations are presented. For simplicity, as with the P10 formulation, the equations are presented assuming a constant smoothing length, in which case they are given by
\begin{align}
\frac{{\rm d}v_a^i}{{\rm d}t} =& - \sum_b m_b \left[ \frac{P_a}{\rho_a^2} + \frac{P_b}{\rho_b^2} \right]\frac{\partial W_{ab}}{\partial x_a^i} \nonumber \\
& + \frac{3}{2 \mu_0} \sum_b m_b  \left[ \frac{B_a^2}{\rho_a^2} + \frac{B_b^2}{\rho_b^2} \right] \frac{\partial W_{ab}}{\partial x_a^i} \nonumber \\
& - \frac{1}{\mu_0} \epsilon_{jkl} \sum_b m_b (A_a^k - A_b^k) \left[ \frac{B_a^j}{\rho_a^2} + \frac{B_b^j}{\rho_b^2} \right] \frac{\partial^2 W_{ab}}{\partial x_a^i \partial x_a^l} \nonumber \\
& - \sum_b m_b \left[ \frac{A_a^i}{\rho_a^2} J_a^k + \frac{A_b^i}{\rho_b^2} J_b^k \right] \frac{\partial W_{ab}}{\partial x_a^k} \nonumber \\
&- \sum_b m_b (A_a^k - A_b^k) \left[ \frac{J_a^k}{\rho_a^2} + \frac{J_b^k}{\rho_b^2} \right] \frac{\partial W_{ab}}{\partial x_a^i} ,
\label{eq:volume-force}
\end{align}
where $J_a^k$ is given by (\ref{eq:symm_J}). 

It is important to recognize that the second term in (\ref{eq:p10-force}) and (\ref{eq:volume-force}) is present irrespective of the reconstruction used for the magnetic field or the time evolution of the vector potential. The third term also arises solely due to the choice of magnetic field reconstruction given by (\ref{eq:b}). This means that using a different formulation for ${\rm d}{\bf A}/{\rm d}t$ can only change the fourth term in (\ref{eq:p10-force}). In the case of the new volume integral formulation, this leads to the fifth term in (\ref{eq:volume-force}), due to the second term in (\ref{eq:integral_dadt_gauge}).

The issues with the conservative equations of motion for the P10 formulation are not solved with this new set of equations. Furthermore, the resultant equations of motion do not seem to yield the correct MHD equations of motion. In Appendix C of \cite{price10}, the translation of (\ref{eq:p10-force}) to the continuum limit was derived by using the basic summation interpolant, $A_a = \sum_b (m_b/\rho_b) A_b W_{ab}$ and its derivative, to understand each term. \cite{price10} demonstrated that (\ref{eq:p10-force}) does indeed solve the equation
\begin{equation}
\frac{	{\rm d}v^i}{{\rm d}t} = - \frac{1}{\rho} \frac{\partial P}{\partial x^i} + \frac{1}{\rho} \left[ J^j \frac{\partial A^j}{\partial x^i} - J^j \frac{\partial A^i}{\partial x^j} \right],
\end{equation}
where the MHD terms derive from ${\bf J} \times {\bf B}$. The fifth term in (\ref{eq:volume-force}), which is the new term stemming from the volume integral formulation, can be straightforwardly shown to be equivalent to $2 (J^j / \rho) (\partial A^j / \partial x^i)$, which means (\ref{eq:volume-force}) solves
\begin{equation}
\frac{	{\rm d}v^i}{{\rm d}t} = - \frac{1}{\rho} \frac{\partial P}{\partial x^i} + \frac{1}{\rho} \left[ 3 J^j \frac{\partial A^j}{\partial x^i} - J^j \frac{\partial A^i}{\partial x^j} \right] . 
\end{equation}
This is a problem. For this reason, the conservative equations of motion are not tested in this work, opting instead for a hybrid approach.

\subsection{Hybrid approach}

A hybrid approach for the equations of motion was found by \cite{price10} to yield the most success. In this case, the magnetic field is reconstructed from the vector potential, which is then used in the standard SPMHD equations of motion. This is the approach followed in this work. Specifically, the momentum equation that is used is given by
\begin{align}
\frac{{\rm d}{\bf v}_a}{{\rm d}t} =& - \sum_b m_b \left[ \frac{P_a}{\Omega_a \rho_a^2} \nabla_a W_{ab}(h_a) + \frac{P_b}{\Omega_b \rho_b^2} \nabla_a W_{ab}(h_b) \right] \nonumber \\
&- \frac{1}{2\mu_0} \sum_b m_b \left[ \frac{B_a^2}{\Omega_a \rho_a^2} \nabla_a W_{ab}(h_a) + \frac{B_b^2}{\Omega_b \rho_b^2}\nabla_a W_{ab}(h_b) \right] \nonumber \\
&+ \frac{1}{\mu_0} \sum_b m_b \frac{({\bf B}_b - {\bf B}_a)}{\Omega_b \rho_b^2} {\bf B}_a \cdot \nabla_a W_{ab}(h_b) .
\end{align}
This includes the stabilisation of the magnetic tension (third term) through subtraction of ${\bf B} (\nabla \cdot {\bf B})$ \cite{bot01}.

\section{Results: 3D Orszag-Tang vortex}
\label{sec:tests}

Tests of the new formulation focus on a 3D version of the Orszag-Tang vortex \cite{ot79, phantom}. This test is particularly notable as a stumbling block for the P10 formulation. It is important to consider a fully 3D test case, since magnetic fields for 2D test problems can be specified solely in terms of a constant $A_z$. Results are compared between the new volume integral formulation, the P10 formulation, and standard SPMHD evolving ${\bf B}$ directly.

\subsection{Numerical setup}

The Orszag-Tang vortex is extended to three dimensions by creating a thin slab. The spatial domain is $x, y \in [-0.5, 0.5]$ and $z \in [0, 3 \sqrt{6} / 64]$. This gives a thickness of the slab of ${\sim}0.1148$, chosen because the particles are arranged on a close-packed triangular lattice. Periodic boundary conditions are used. The initial density and pressure are uniform, with $\rho = 25 / (36 \pi)$, $P = 5 / (12 \pi)$ and $\gamma = 5/3$. The initial velocity field is $[v_x, v_y, v_z] = [- \sin(2\pi y), \sin(2 \pi x), 0]$, and initial magnetic field $[B_x, B_y, B_z] = [- B_0 \sin(2 \pi y), B_0 \sin(4 \pi x), 0]$, with $B_0 = 1 / \sqrt{4 \pi}$. The corresponding vector potential is $A_z = B_0 / (2 \pi) [\cos(2 \pi y) + \cos(4 \pi x) / 2 ]$. 

The lowest resolution calculation uses a total of 42~624 particles arranged initially on a close-packed triangular lattice, with $[n_{\rm x}, n_{\rm y}, n_{\rm z}] = [64, 74, 9]$, and the highest resolution calculation uses 2~727~936 particles with $[n_{\rm x}, n_{\rm y}, n_{\rm z}]=[256, 296, 36]$.  Resolutions are tested for $n_{\rm x} \in [64, 128, 256]$. Note that $n_{\rm z}$ also doubles with each step increase in resolution so that the slab retains the same physical thickness.

\subsection{Qualitative analysis}

Fig.~\ref{fig:density} shows density slices through the midplane for all calculations. Both the new vector potential implementation derived from volume integrals and the P10 vector potential formulation experience numerical instability, with large low-density voids appearing in the obtained solutions. These appear earlier for higher numerical resolution, occurring at $t{\sim}0.5$ for the $n_{\rm x}=64$ calculations, $t{\sim}0.29$ for the $n_{\rm x}=128$ calculations, and $t{\sim}0.2$ for the $n_{\rm x}=256$ calculations. Reference solutions are provided by calculations that directly evolve the magnetic field. 

That the new vector potential discretisation based on volume integrals experience numerical instability in the same manner and with the same timing as the P10 formulation suggests that the extra term in (\ref{eq:integral_dadt_gauge}) has negligible effect.

One important observation is that, aside from the region where the numerical instability occurs, the solutions obtained with the vector potential formulations are in agreement with the solution obtained when the magnetic field is directly evolved. Thus, using the vector potential in SPMHD does indeed represent a valid prescription for the magnetic field. All that is required is for the vector potential to be implemented in a numerically stable manner.

\subsection{Magnetic energy}

\begin{figure}
\includegraphics[width=\columnwidth]{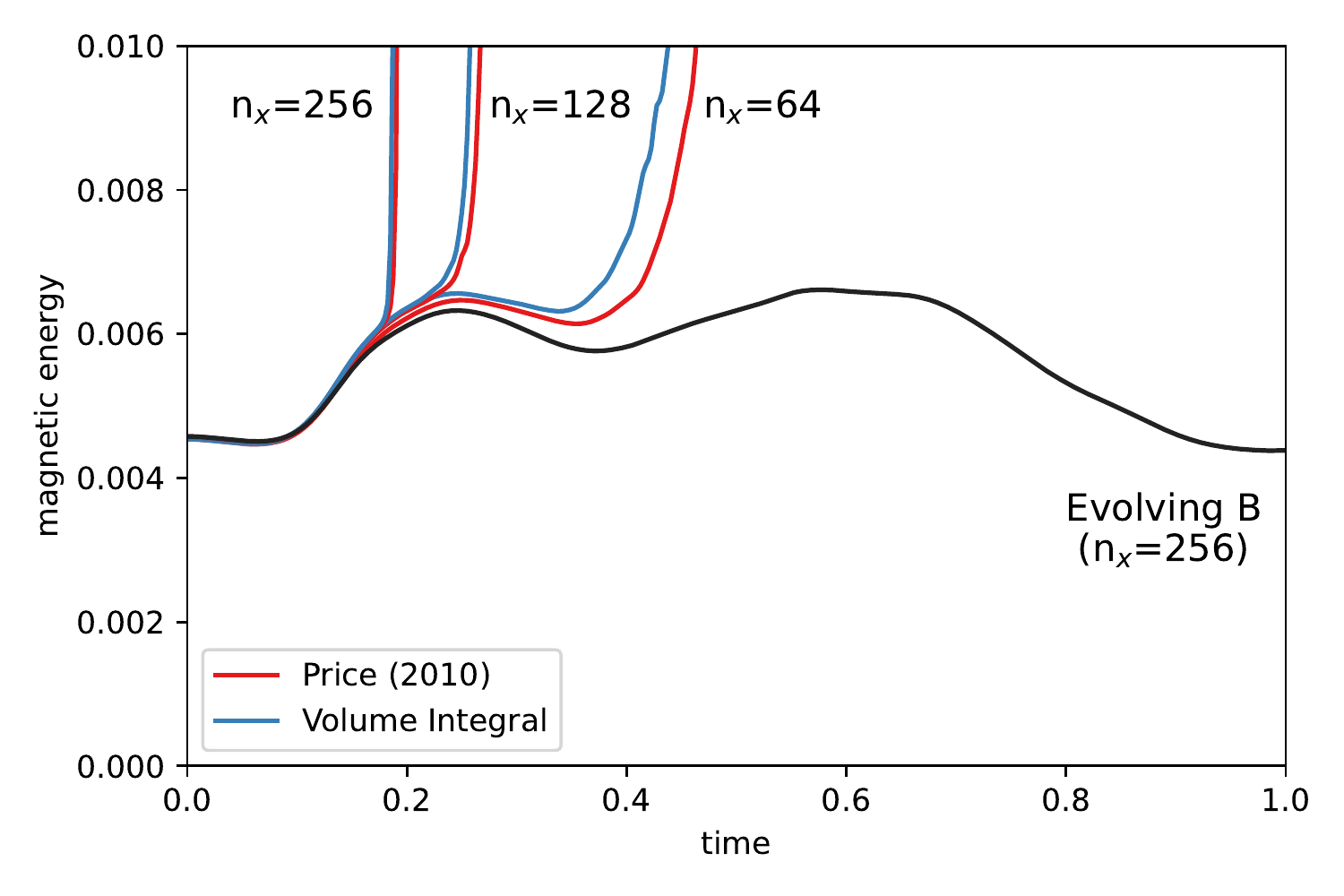}
\caption{Time evolution of the magnetic energy for the new volume integral formulation (blue) and the P10 formulation (red) for resolutions of $n_{\rm x}=64$, 128 and 256 particles. The magnetic energy for direct evolution of the magnetic field for $n_{\rm x}=256$ is shown for comparison (black).}
\label{fig:be}
\end{figure}
Fig.~\ref{fig:be} shows the time evolution of the magnetic energy. The new discretisation of the vector potential based on volume integrals experiences the same exponential growth of magnetic energy as the P10 formulation. The onset of numerical instability occurs earlier as the numerical resolution increases, occurring approximately at the same time as the P10 formulation. This suggests the extra term in (\ref{eq:integral_dadt_gauge}) has negligible effect.

\subsection{Components of the vector potential}

\begin{figure}
\includegraphics[width=\columnwidth]{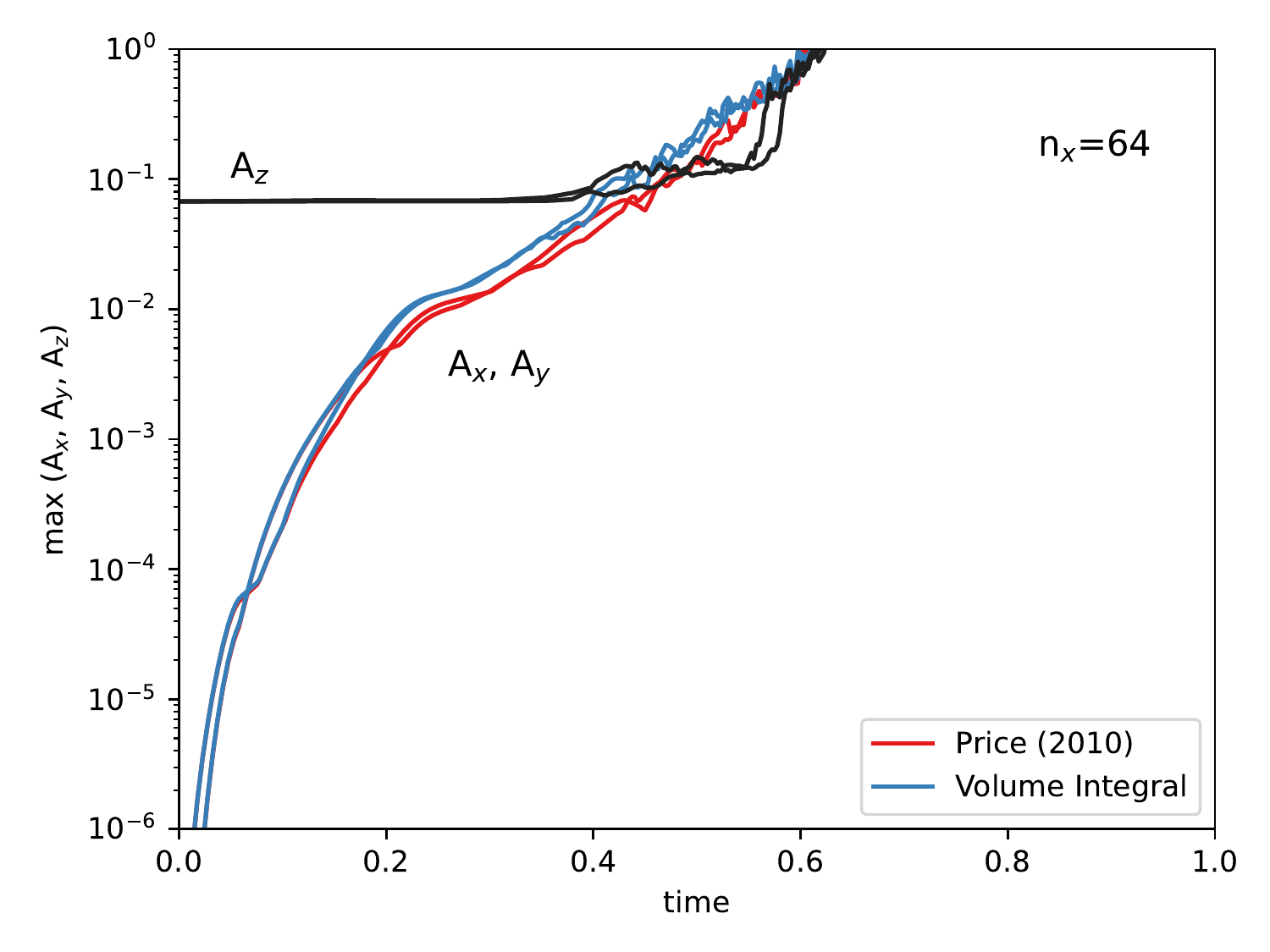}\\
\includegraphics[width=\columnwidth]{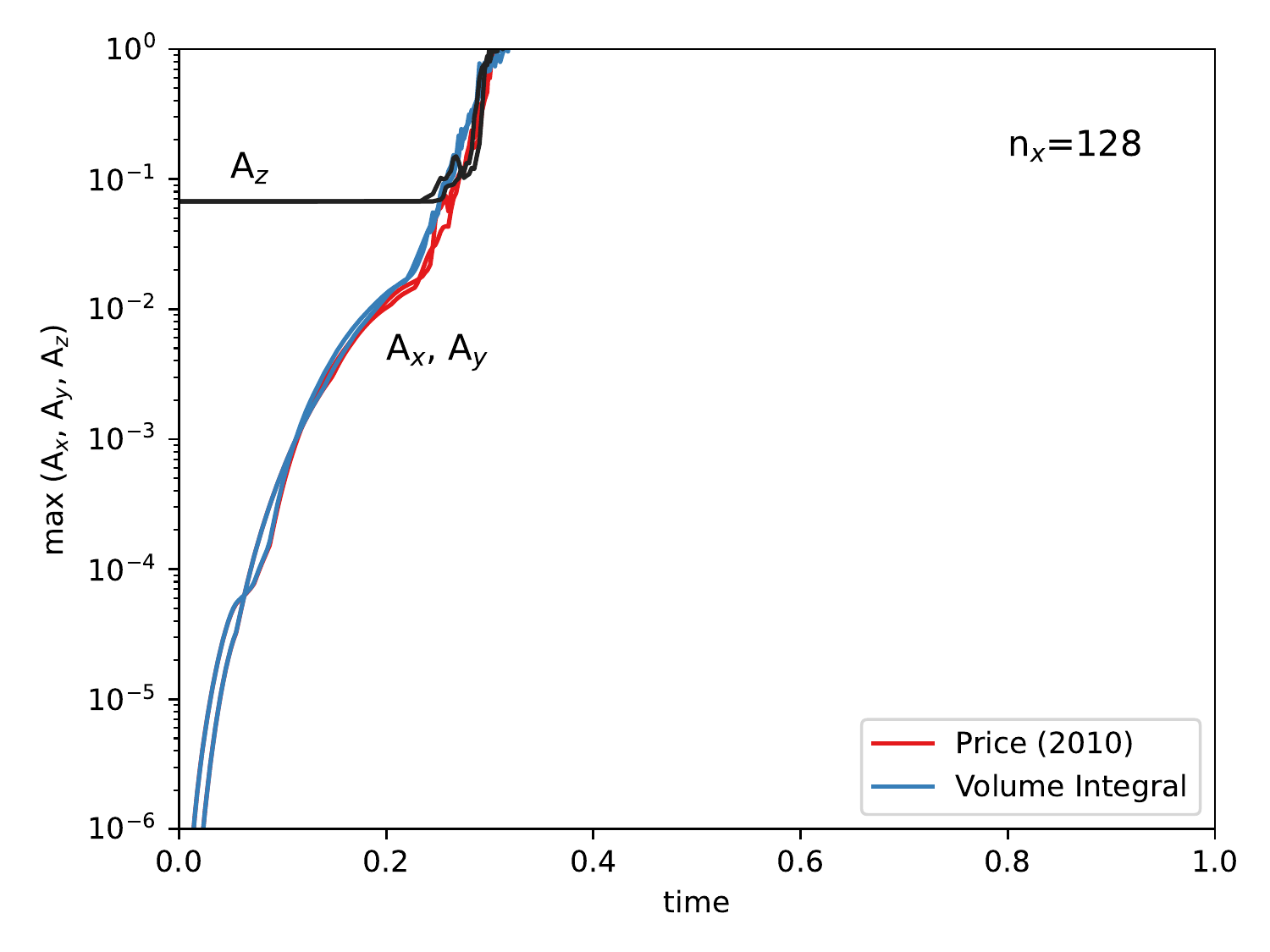}\\
\includegraphics[width=\columnwidth]{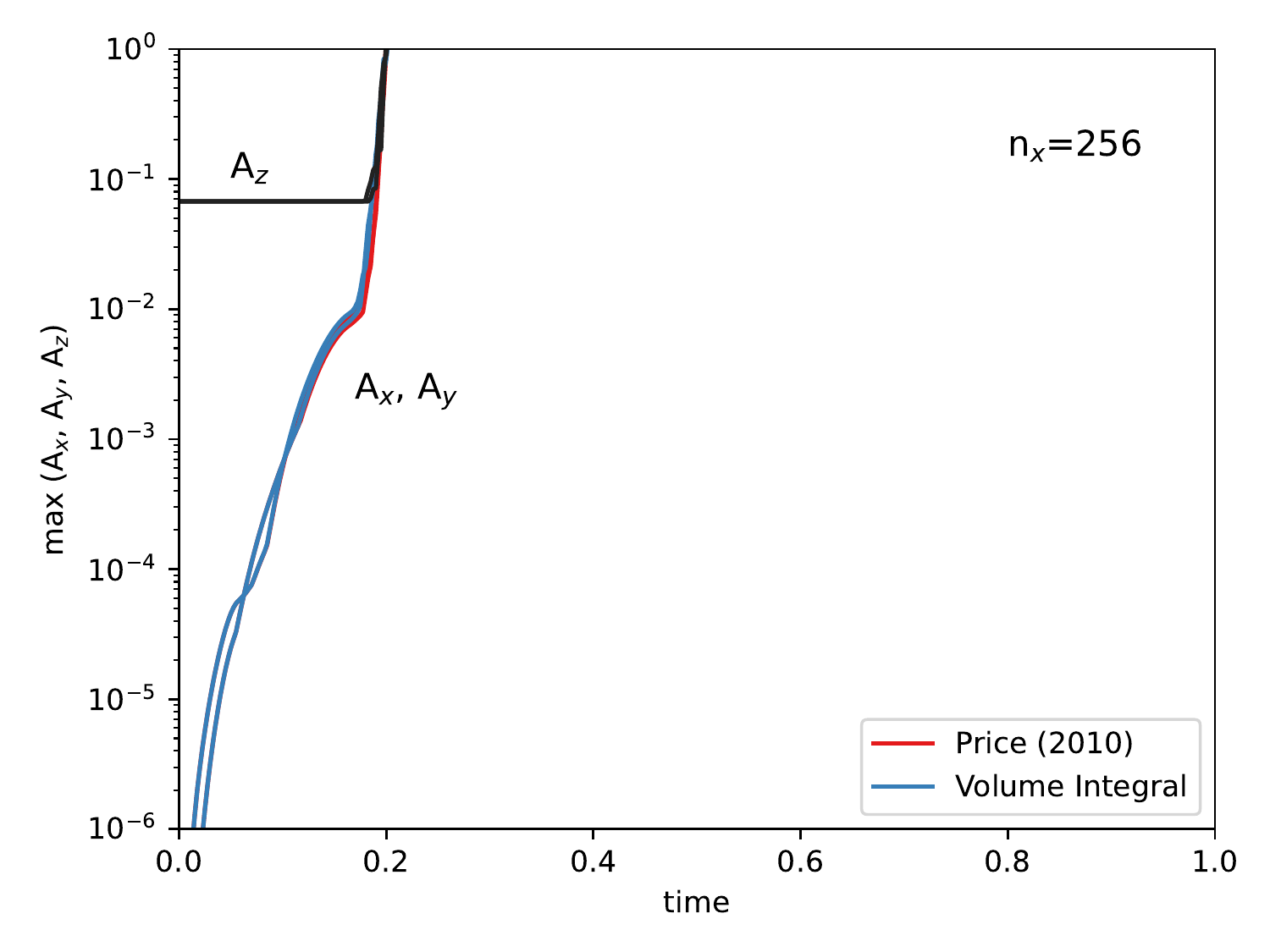}
\caption{Time evolution of the maximum value of the vector potential components for the new volume integral formulation (blue) and the formulation by \cite{price10} (red) for resolutions of $n_{\rm x}=64$, 128 and 256 particles (top to bottom panels). In all cases, numerical instability occurs once the $A_x$ and $A_y$ components, which are initially zero, become comparable to the $A_z$ component.}
\label{fig:a}
\end{figure}

Fig.~\ref{fig:a} shows the time evolution of the components of the vector potential for the $n_{\rm x}=64$, 128 and 256 calculations. The initially zero $A_x$ and $A_y$ components of the vector potential undergo exponential amplification. Once $A_x$ and $A_y$ reach parity with the $A_z$ component, then all three components undergo exponential growth. There is negligible difference between the volume integral and P10 formulations. The initial tracks for the $A_x$ and $A_z$ components are common between the three resolutions, up to $t{\sim}0.2$, at which time the particles break off lattice.

\subsection{$z$ component of the magnetic field}

\begin{figure}
\includegraphics[width=\columnwidth]{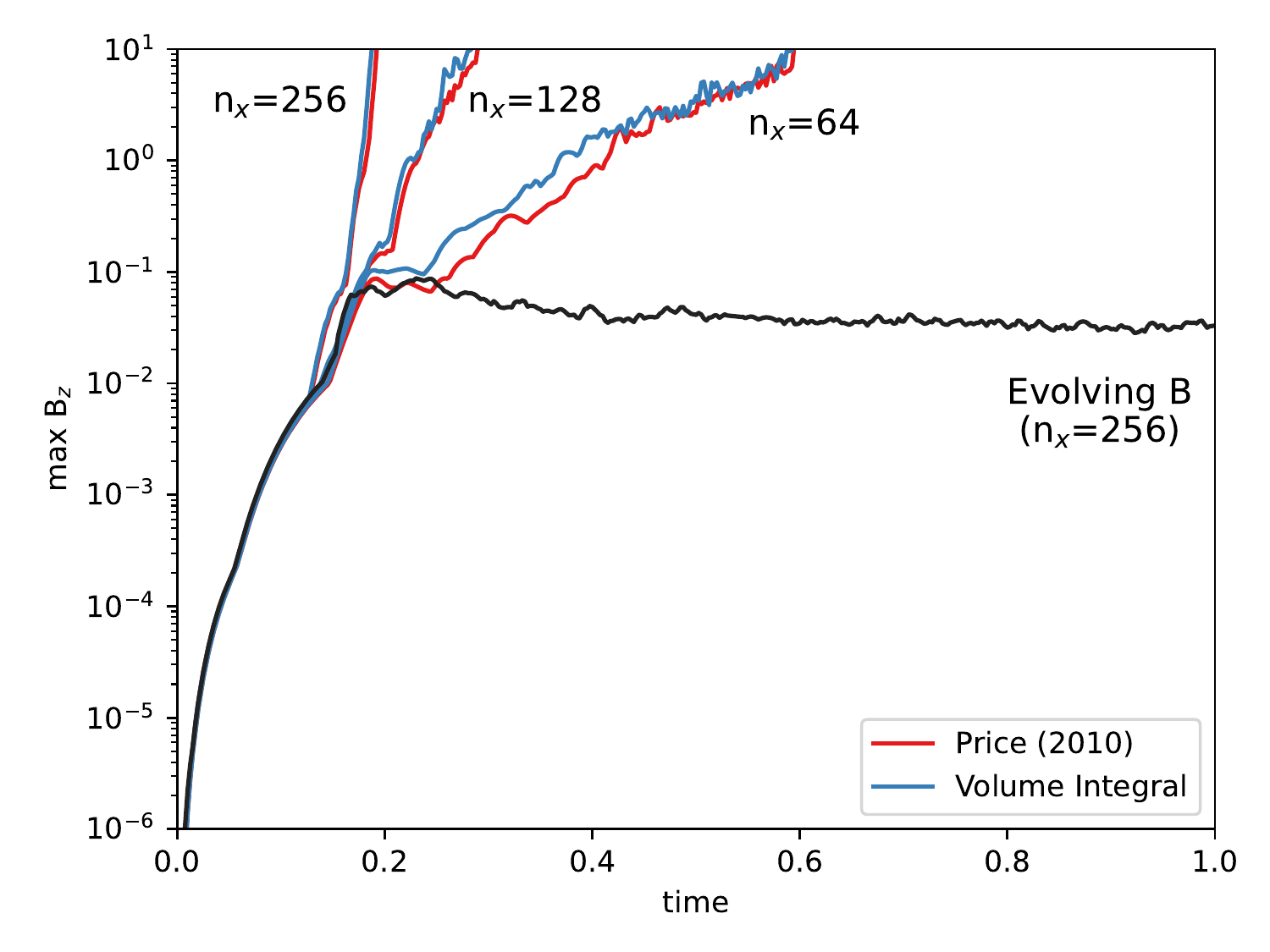}
\caption{Time evolution of the maximum $B_z$ over all particles. The $B_z$ component exhibits nearly identical exponential growth for all vector potential calculations and the calculation directly evolving the magnetic field up to $t{\sim}0.2$. At this time, the particles break off lattice. When directly evolving the magnetic field, $B_z$ saturates at ${\sim}0.1$, whereas the vector potential calculations continue to experience exponential growth. }
\label{fig:bz}
\end{figure}

Fig.~\ref{fig:bz} shows the time evolution of the $z$ component of the magnetic field. The vector potential calculations, irrespective of resolution, and the calculation directly evolving the magnetic field all show the same characteristic initial exponential growth of $B_z$. The calculations diverge at $t{\sim}0.2$. At this point, $B_z$ saturates for the calculation directly evolving ${\bf B}$. Conversely, the vector potential calculations continue to display exponential growth, with faster growth rates as the numerical resolution increases. There is negligible difference between the volume integral and P10 formulations.

\begin{figure}
\includegraphics[width=\columnwidth]{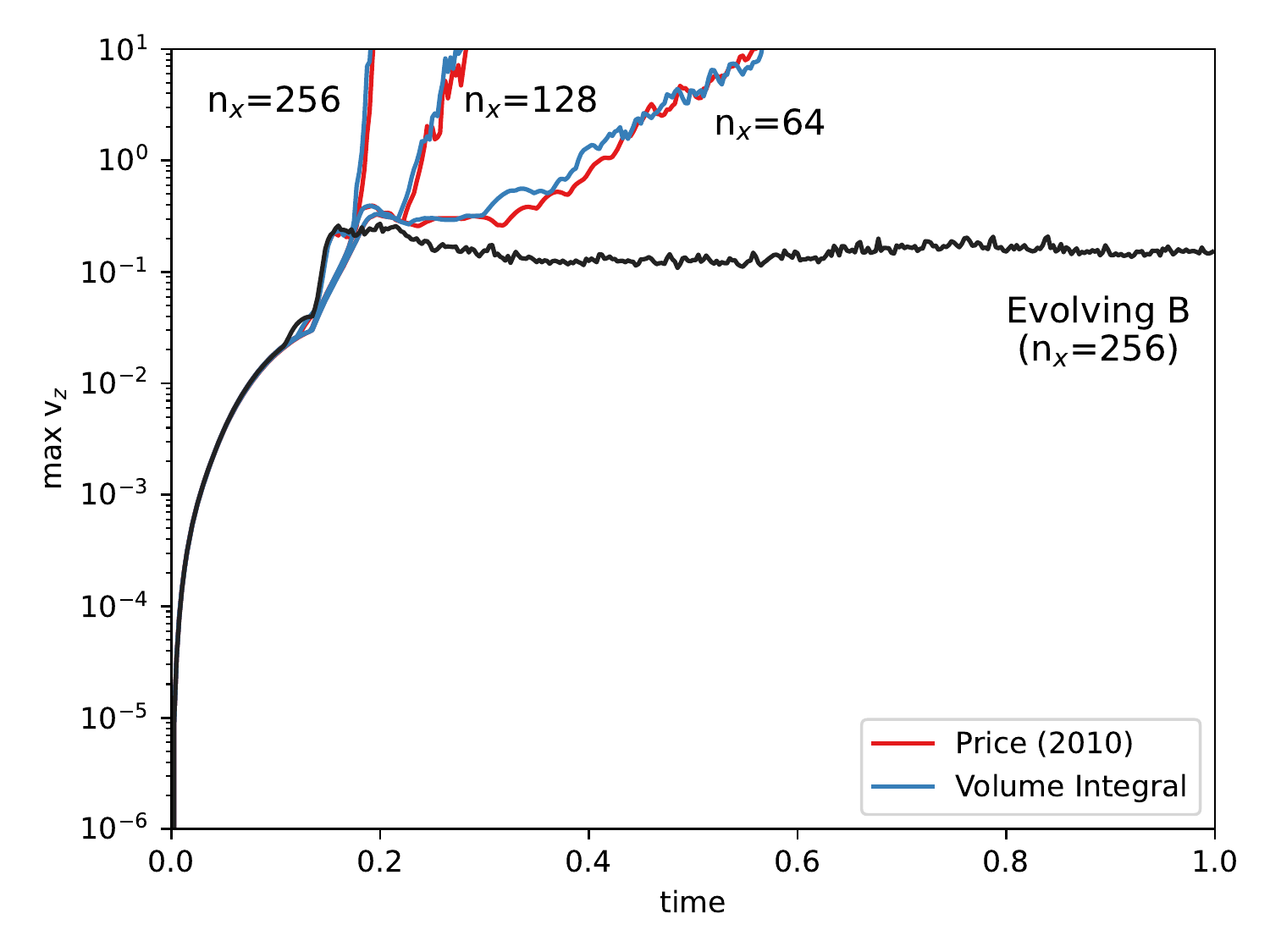}
\caption{Time evolution of the maximum $v_z$ over all particles. The correlation with $B_z$ is apparent (see Fig.~\ref{fig:bz}). When directly evolving the magnetic field, $v_z$ has initial exponential growth, but then saturates. For the vector potential calculations, $v_z$ continues to grow exponentially.}
\label{fig:vz}
\end{figure}

Fig.~\ref{fig:vz} shows the time evolution of $v_z$. From this, it becomes clear that the growth of $B_z$ is due to the growth of $v_z$, which is a consequence of performing these calculations in 3D. In 2D calculations, $v_z$ remains zero, but this is not the case in 3D.

\section{Conclusion}
\label{sec:conclusions}

A novel formulation for evolving the vector potential in SPMHD has been derived. By considering the induction equation in integral form, an altered form of the discretised `reverse advection' term is obtained. In particular, this reverse advection term uses an interpolation of the velocity of neighbouring particles instead of a particle's velocity directly. A Galilean invariant discretisation can be obtained through suitable gauge choice ($\phi = - {\bf v} \cdot {\bf A}$). The Galilean invariant discretisation is equivalent to the P10 formulation with an additional term.

The new discretisation is tested using a 3D version of the Orszag-Tang vortex. This test has proven difficult to stably simulate for prior SPMHD formulations of the vector potential \cite{price10}. Results are compared to the P10 formulation and to a calculation where the magnetic field is evolved directly (not using the vector potential). 

The key result is that the volume integral discretisation exhibits the same numerical instability as the P10 formulation. Initially, calculations proceed smoothly, but, at some later time, the magnetic energy increases exponentially. The onset of instability occurs earlier for higher numerical resolutions. This appears to be triggered by particles breaking off lattice at $t{\sim}0.2$, and, more concretely, once the initially zero $A_x$ and $A_y$ components reach parity with $A_z$. 

A solution to the numerical instability is, in theory, straightforward -- use the conservative equations of motion as derived from the Lagrangian. If the magnetic energy (and hence total energy) is unphysically increasing, then using a discretisation that conserves total energy should fix this. This does not work in practice. The conservative equations of motion for the P10 formulation are plagued by negative pressure and a noisy direct second derivative of the kernel. These particular problems do not originate from the evolution equation used for the vector potential evolution equation, and are thus present in the conservative equations of motion derived for the new volume integral formulation. The questions on how to solve these problems remain open.

Even more troubling is that the conservative equations of motion for the volume integral formulation do not appear to actually solve the MHD equations. It was demonstrated in \cite{price10} that the P10 equations of motion do indeed translate to the MHD equations in the continuum limit. The equations of motion for the new volume integral formulation encapsulate the P10 equations of motion plus an additional, non-zero term. They therefore cannot be equivalent to the MHD equations.

While the exploration of induction equation expressed in terms of a volume integral has led to a novel discretisation of the vector potential in SPMHD, it suffers from the same shortcomings of the P10 formulation. The search continues for a stable formulation of SPMHD that guarantees a divergence-free magnetic field.

\section*{Acknowledgment}

Figures were made and analysis conducted using {\sc Sarracen} \cite{sarracen}, a Python-based visualization and analysis package for SPH data. This research was enabled in part by advanced computing resources provided by the Digital Research Alliance of Canada, the organization responsible for digital research infrastructure in Canada, and ACENET, the regional partner in Atlantic Canada.



\bibliographystyle{IEEEtran.bst}
\bibliography{IEEEabrv,bib}
\end{document}